\documentclass{article}

    \usepackage{graphicx} 
    \usepackage{adjustbox} 
    \usepackage{color} 
    \usepackage{enumerate} 
    \usepackage{geometry} 
    \usepackage{amsmath} 
    \usepackage{amssymb} 
    \usepackage[mathletters]{ucs} 
    \usepackage[utf8x]{inputenc} 
    \usepackage{fancyvrb} 
    \usepackage{grffile} 
    \usepackage{hyperref}
    \usepackage{authblk}
    \usepackage{longtable} 
    \usepackage{booktabs}  
    
    \author[1]{Jacopo Corbetta\thanks{jacorbetta@zeliade.com, Corresponding author}}
    \author[1]{Pierre Cohort\thanks{pcohort@zeliade.com}}
    \author[1]{Ismail Laachir\thanks{ilaachir@zeliade.com}}
    \author[1]{Claude Martini\thanks{cmartini@zeliade.com}}
      
    \affil[1]{Zeliade Systems, 56 rue Jean-Jacques Rousseau, Paris, France}

    \definecolor{orange}{cmyk}{0,0.4,0.8,0.2}
    \definecolor{darkorange}{rgb}{.71,0.21,0.01}
    \definecolor{darkgreen}{rgb}{.12,.54,.11}
    \definecolor{myteal}{rgb}{.26, .44, .56}
    \definecolor{gray}{gray}{0.45}
    \definecolor{lightgray}{gray}{.95}
    \definecolor{mediumgray}{gray}{.8}
    \definecolor{inputbackground}{rgb}{.95, .95, .85}
    \definecolor{outputbackground}{rgb}{.95, .95, .95}
    \definecolor{traceback}{rgb}{1, .95, .95}
    \definecolor{red}{rgb}{.6,0,0}
    \definecolor{green}{rgb}{0,.65,0}
    \definecolor{brown}{rgb}{0.6,0.6,0}
    \definecolor{blue}{rgb}{0,.145,.698}
    \definecolor{purple}{rgb}{.698,.145,.698}
    \definecolor{cyan}{rgb}{0,.698,.698}
    \definecolor{lightgray}{gray}{0.5}
    
    \definecolor{darkgray}{gray}{0.25}
    \definecolor{lightred}{rgb}{1.0,0.39,0.28}
    \definecolor{lightgreen}{rgb}{0.48,0.99,0.0}
    \definecolor{lightblue}{rgb}{0.53,0.81,0.92}
    \definecolor{lightpurple}{rgb}{0.87,0.63,0.87}
    \definecolor{lightcyan}{rgb}{0.5,1.0,0.83}
    
    \DefineVerbatimEnvironment{Highlighting}{Verbatim}{commandchars=\\\{\}}

    \title{Robust calibration and arbitrage-free interpolation of SSVI slices}

\makeatletter
\def\PY@reset{\let\PY@it=\relax \let\PY@bf=\relax%
    \let\PY@ul=\relax \let\PY@tc=\relax%
    \let\PY@bc=\relax \let\PY@ff=\relax}
\def\PY@tok#1{\csname PY@tok@#1\endcsname}
\def\PY@toks#1+{\ifx\relax#1\empty\else%
    \PY@tok{#1}\expandafter\PY@toks\fi}
\def\PY@do#1{\PY@bc{\PY@tc{\PY@ul{%
    \PY@it{\PY@bf{\PY@ff{#1}}}}}}}
\def\PY#1#2{\PY@reset\PY@toks#1+\relax+\PY@do{#2}}

\expandafter\def\csname PY@tok@gd\endcsname{\def\PY@tc##1{\textcolor[rgb]{0.63,0.00,0.00}{##1}}}
\expandafter\def\csname PY@tok@gu\endcsname{\let\PY@bf=\textbf\def\PY@tc##1{\textcolor[rgb]{0.50,0.00,0.50}{##1}}}
\expandafter\def\csname PY@tok@gt\endcsname{\def\PY@tc##1{\textcolor[rgb]{0.00,0.27,0.87}{##1}}}
\expandafter\def\csname PY@tok@gs\endcsname{\let\PY@bf=\textbf}
\expandafter\def\csname PY@tok@gr\endcsname{\def\PY@tc##1{\textcolor[rgb]{1.00,0.00,0.00}{##1}}}
\expandafter\def\csname PY@tok@cm\endcsname{\let\PY@it=\textit\def\PY@tc##1{\textcolor[rgb]{0.25,0.50,0.50}{##1}}}
\expandafter\def\csname PY@tok@vg\endcsname{\def\PY@tc##1{\textcolor[rgb]{0.10,0.09,0.49}{##1}}}
\expandafter\def\csname PY@tok@m\endcsname{\def\PY@tc##1{\textcolor[rgb]{0.40,0.40,0.40}{##1}}}
\expandafter\def\csname PY@tok@mh\endcsname{\def\PY@tc##1{\textcolor[rgb]{0.40,0.40,0.40}{##1}}}
\expandafter\def\csname PY@tok@go\endcsname{\def\PY@tc##1{\textcolor[rgb]{0.53,0.53,0.53}{##1}}}
\expandafter\def\csname PY@tok@ge\endcsname{\let\PY@it=\textit}
\expandafter\def\csname PY@tok@vc\endcsname{\def\PY@tc##1{\textcolor[rgb]{0.10,0.09,0.49}{##1}}}
\expandafter\def\csname PY@tok@il\endcsname{\def\PY@tc##1{\textcolor[rgb]{0.40,0.40,0.40}{##1}}}
\expandafter\def\csname PY@tok@cs\endcsname{\let\PY@it=\textit\def\PY@tc##1{\textcolor[rgb]{0.25,0.50,0.50}{##1}}}
\expandafter\def\csname PY@tok@cp\endcsname{\def\PY@tc##1{\textcolor[rgb]{0.74,0.48,0.00}{##1}}}
\expandafter\def\csname PY@tok@gi\endcsname{\def\PY@tc##1{\textcolor[rgb]{0.00,0.63,0.00}{##1}}}
\expandafter\def\csname PY@tok@gh\endcsname{\let\PY@bf=\textbf\def\PY@tc##1{\textcolor[rgb]{0.00,0.00,0.50}{##1}}}
\expandafter\def\csname PY@tok@ni\endcsname{\let\PY@bf=\textbf\def\PY@tc##1{\textcolor[rgb]{0.60,0.60,0.60}{##1}}}
\expandafter\def\csname PY@tok@nl\endcsname{\def\PY@tc##1{\textcolor[rgb]{0.63,0.63,0.00}{##1}}}
\expandafter\def\csname PY@tok@nn\endcsname{\let\PY@bf=\textbf\def\PY@tc##1{\textcolor[rgb]{0.00,0.00,1.00}{##1}}}
\expandafter\def\csname PY@tok@no\endcsname{\def\PY@tc##1{\textcolor[rgb]{0.53,0.00,0.00}{##1}}}
\expandafter\def\csname PY@tok@na\endcsname{\def\PY@tc##1{\textcolor[rgb]{0.49,0.56,0.16}{##1}}}
\expandafter\def\csname PY@tok@nb\endcsname{\def\PY@tc##1{\textcolor[rgb]{0.00,0.50,0.00}{##1}}}
\expandafter\def\csname PY@tok@nc\endcsname{\let\PY@bf=\textbf\def\PY@tc##1{\textcolor[rgb]{0.00,0.00,1.00}{##1}}}
\expandafter\def\csname PY@tok@nd\endcsname{\def\PY@tc##1{\textcolor[rgb]{0.67,0.13,1.00}{##1}}}
\expandafter\def\csname PY@tok@ne\endcsname{\let\PY@bf=\textbf\def\PY@tc##1{\textcolor[rgb]{0.82,0.25,0.23}{##1}}}
\expandafter\def\csname PY@tok@nf\endcsname{\def\PY@tc##1{\textcolor[rgb]{0.00,0.00,1.00}{##1}}}
\expandafter\def\csname PY@tok@si\endcsname{\let\PY@bf=\textbf\def\PY@tc##1{\textcolor[rgb]{0.73,0.40,0.53}{##1}}}
\expandafter\def\csname PY@tok@s2\endcsname{\def\PY@tc##1{\textcolor[rgb]{0.73,0.13,0.13}{##1}}}
\expandafter\def\csname PY@tok@vi\endcsname{\def\PY@tc##1{\textcolor[rgb]{0.10,0.09,0.49}{##1}}}
\expandafter\def\csname PY@tok@nt\endcsname{\let\PY@bf=\textbf\def\PY@tc##1{\textcolor[rgb]{0.00,0.50,0.00}{##1}}}
\expandafter\def\csname PY@tok@nv\endcsname{\def\PY@tc##1{\textcolor[rgb]{0.10,0.09,0.49}{##1}}}
\expandafter\def\csname PY@tok@s1\endcsname{\def\PY@tc##1{\textcolor[rgb]{0.73,0.13,0.13}{##1}}}
\expandafter\def\csname PY@tok@kd\endcsname{\let\PY@bf=\textbf\def\PY@tc##1{\textcolor[rgb]{0.00,0.50,0.00}{##1}}}
\expandafter\def\csname PY@tok@sh\endcsname{\def\PY@tc##1{\textcolor[rgb]{0.73,0.13,0.13}{##1}}}
\expandafter\def\csname PY@tok@sc\endcsname{\def\PY@tc##1{\textcolor[rgb]{0.73,0.13,0.13}{##1}}}
\expandafter\def\csname PY@tok@sx\endcsname{\def\PY@tc##1{\textcolor[rgb]{0.00,0.50,0.00}{##1}}}
\expandafter\def\csname PY@tok@bp\endcsname{\def\PY@tc##1{\textcolor[rgb]{0.00,0.50,0.00}{##1}}}
\expandafter\def\csname PY@tok@c1\endcsname{\let\PY@it=\textit\def\PY@tc##1{\textcolor[rgb]{0.25,0.50,0.50}{##1}}}
\expandafter\def\csname PY@tok@kc\endcsname{\let\PY@bf=\textbf\def\PY@tc##1{\textcolor[rgb]{0.00,0.50,0.00}{##1}}}
\expandafter\def\csname PY@tok@c\endcsname{\let\PY@it=\textit\def\PY@tc##1{\textcolor[rgb]{0.25,0.50,0.50}{##1}}}
\expandafter\def\csname PY@tok@mf\endcsname{\def\PY@tc##1{\textcolor[rgb]{0.40,0.40,0.40}{##1}}}
\expandafter\def\csname PY@tok@err\endcsname{\def\PY@bc##1{\setlength{\fboxsep}{0pt}\fcolorbox[rgb]{1.00,0.00,0.00}{1,1,1}{\strut ##1}}}
\expandafter\def\csname PY@tok@mb\endcsname{\def\PY@tc##1{\textcolor[rgb]{0.40,0.40,0.40}{##1}}}
\expandafter\def\csname PY@tok@ss\endcsname{\def\PY@tc##1{\textcolor[rgb]{0.10,0.09,0.49}{##1}}}
\expandafter\def\csname PY@tok@sr\endcsname{\def\PY@tc##1{\textcolor[rgb]{0.73,0.40,0.53}{##1}}}
\expandafter\def\csname PY@tok@mo\endcsname{\def\PY@tc##1{\textcolor[rgb]{0.40,0.40,0.40}{##1}}}
\expandafter\def\csname PY@tok@kn\endcsname{\let\PY@bf=\textbf\def\PY@tc##1{\textcolor[rgb]{0.00,0.50,0.00}{##1}}}
\expandafter\def\csname PY@tok@mi\endcsname{\def\PY@tc##1{\textcolor[rgb]{0.40,0.40,0.40}{##1}}}
\expandafter\def\csname PY@tok@gp\endcsname{\let\PY@bf=\textbf\def\PY@tc##1{\textcolor[rgb]{0.00,0.00,0.50}{##1}}}
\expandafter\def\csname PY@tok@o\endcsname{\def\PY@tc##1{\textcolor[rgb]{0.40,0.40,0.40}{##1}}}
\expandafter\def\csname PY@tok@kr\endcsname{\let\PY@bf=\textbf\def\PY@tc##1{\textcolor[rgb]{0.00,0.50,0.00}{##1}}}
\expandafter\def\csname PY@tok@s\endcsname{\def\PY@tc##1{\textcolor[rgb]{0.73,0.13,0.13}{##1}}}
\expandafter\def\csname PY@tok@kp\endcsname{\def\PY@tc##1{\textcolor[rgb]{0.00,0.50,0.00}{##1}}}
\expandafter\def\csname PY@tok@w\endcsname{\def\PY@tc##1{\textcolor[rgb]{0.73,0.73,0.73}{##1}}}
\expandafter\def\csname PY@tok@kt\endcsname{\def\PY@tc##1{\textcolor[rgb]{0.69,0.00,0.25}{##1}}}
\expandafter\def\csname PY@tok@ow\endcsname{\let\PY@bf=\textbf\def\PY@tc##1{\textcolor[rgb]{0.67,0.13,1.00}{##1}}}
\expandafter\def\csname PY@tok@sb\endcsname{\def\PY@tc##1{\textcolor[rgb]{0.73,0.13,0.13}{##1}}}
\expandafter\def\csname PY@tok@k\endcsname{\let\PY@bf=\textbf\def\PY@tc##1{\textcolor[rgb]{0.00,0.50,0.00}{##1}}}
\expandafter\def\csname PY@tok@se\endcsname{\let\PY@bf=\textbf\def\PY@tc##1{\textcolor[rgb]{0.73,0.40,0.13}{##1}}}
\expandafter\def\csname PY@tok@sd\endcsname{\let\PY@it=\textit\def\PY@tc##1{\textcolor[rgb]{0.73,0.13,0.13}{##1}}}

\makeatother

    \definecolor{incolor}{rgb}{0.0, 0.0, 0.5}
    \definecolor{outcolor}{rgb}{0.545, 0.0, 0.0}

    \hypersetup{
      breaklinks=true,  
      colorlinks=true,
      urlcolor=blue,
      linkcolor=darkorange,
      citecolor=darkgreen,
      }
    
\begin{document}

     \hypersetup{ breaklinks=true,  
      colorlinks=true,
      urlcolor=blue,
      linkcolor=darkorange,
      citecolor=darkgreen,      
      }

\begin{figure}
\hfill\includegraphics[height=0.1\textwidth]{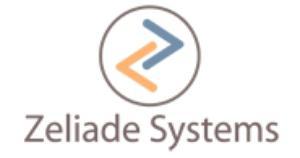}
\end{figure}
\maketitle

	We describe a robust calibration algorithm of a set of SSVI maturity
slices (i.e.~a set of 3 SSVI parameters $\theta_t, \rho_t, \varphi_t$
attached to each option maturity $t$ available on the market), which
grants that these slices are free of Butterfly and of Calendar-Spread
arbitrage. Given such a set of consistent SSVI parameters, we show that
the most natural interpolation/extrapolation \emph{of the parameters}
provides a full continuous volatility surface free of arbitrage. The
numerical implementation is straightforward, robust and quick, yielding
an effective and parsimonious solution to the smile problem, which has
the potential to become a benchmark one.

We thank Antoine Jacquier and Stefano De Marco for useful discussions
and remarks. All remaining errors are ours.

\section{Introduction: the xSSVI
family}\label{introduction-the-xssvi-family}

	Gatheral initiated in 2004 the \emph{Stochastic Volatility Inspired}
parametric model for the volatility smile, a 5-parameter formula for the
total variance at a given maturity. Gatheral's inspiration was of
geometric nature, in relation to Roger Lee's moment formula and also to
his experience of smiles produced by stochastic volatility models like
Heston. SVI fits remarkably well in practice, and it is even difficult
to find circumstances in which SVI fails (Fabien Le Floch provides such
an example on his blog \cite{LeFlochSVI}). Despite its simplicity, the
calibration of SVI is not straightforward, and Zeliade has a whitepaper
with a re-parameterization trick which robustifies a lot the process
(cf. \cite{de2009quasi}; more detailed calculations are available in
Stefano De Marco's PhD Thesis, \cite{de2010probability}).

SVI misses 2 important features: it does not model the whole volatility
surface, and there are no known conditions on SVI parameters which grant
absence of arbitrage (even tractable sufficient conditions).

Then comes SSVI: many teams worked on producing an SVI-like model for
the whole volatility surface in years around 2010, and the only
successful one was the Jim Gatheral and Antoine Jacquier pair, who
designed the \emph{Surface SVI} model which has the 2 features SVI
missed (cf. \cite{GatheralJacquier}). SSVI is (this may seem natural)
parameterized by the ATM (Forward) total variance curve $\theta_t$, so
it will automatically fit perfectly the ATMF point, a \emph{constant}
correlation parameter $\rho$ (which should play the role of the leverage
parameter), and a curvature curve $\varphi$:

\begin{equation}\label{eq: SSVI}
w(k,\theta_t) = \frac{\theta_t}{2}\left( 1  + \rho \varphi(\theta_t) k + \sqrt{(\varphi(\theta_t)k + \rho)^2 + (1-\rho^2)}\right)
\end{equation}

Each smile has so only 3 parameters, and explicit and tractable
\emph{sufficient} conditions have been obtained by Gatheral and Jacquier
to preclude Butterfly arbitrage:

\begin{equation}\label{eq: buttSuff1}
\theta_t \varphi(\theta_t) \leq \frac{4}{1+|\rho|}
\end{equation}

\begin{equation}\label{eq: buttSuff2}
\theta_t \varphi(\theta_t)^2 \leq \frac{4}{1+|\rho|}
\end{equation}

	Moreover, necessary and sufficient no Calendar Spread conditions are
provided. SSVI works reasonably well in practice, and its calibration is
easier than SVI. Yet the fact to keep the correlation constant across
maturities depreciates the fit quality. Sebas Hendriks (student of Kees
Osterlee at the University of Delft, during his master internship at
Zeliade) and Claude Martini tackled this issue (cf.
\cite{hendriks2017extended}): they managed to obtain simple necessary
and sufficient conditions for the consistency of 2 SSVI \emph{slices}
attached to different maturities; such conditions are not available for
SVI smiles. By a SSVI slice we mean the SSVI parameterization at a given
maturity, with its own correlation parameter, possibly depending on the
maturity. Conditions for the absence of Calendar Spread arbitrage in
continuous time follow, and the corresponding \emph{extended SSVI}
(eSSVI) model extends SSVI with a maturity-dependent $\rho(\theta_t)$.
An explicit representation formula for the correlation is also obtained,
which allows to produce easily concrete low-dimensional parameterization
for the correlation curve. Some power-law type examples are provided.

	The effective calibration of eSSVI has not been investigated yet, this
is the purpose of this note. We proceed in 2 steps, each of them having
some interest on its own:

\begin{enumerate}
\def\labelenumi{\arabic{enumi}.}
\itemsep1pt\parskip0pt\parsep0pt
\item
  We calibrate SSVI (equivalently, eSSVI) slices to the available
  maturities on the market in a way which grants the absence of
  Butterfly and of Calendar Spread arbitrage, making use of a very
  robust calibration algorithm, which \emph{does not use any black box
  optimizer} beyond a one-dimensional Brent algorithm.
\item
  We show that the most naive interpolation/extrapolation scheme of the
  \emph{slice parameters} is arbitrage free. This is an unexpected and
  remarkable property of eSSVI.
\end{enumerate}

We obtain therefore a continuous time arbitrage-free eSSVI model
calibrated to the market. We discuss in the conclusion the virtues of
this scheme, that we consider as the quickest and cheapest way (so far)
to solve (not perfectly though, but with a sufficient accuracy in many
situations but the most demanding ones) the smile problem.

	Notice that the eSSVI approach is \emph{model-free} in the sense that it
does not start from specifying a dynamic of the underlying, and then
compute an arbitrage-free price and eventually the corresponding implied
volatility surface: it directly tackles the implied volatility surface.
In this sense, results in Lee (cf. \cite{lee2005implied}) can not be
applied directly; those type of results have been originally inspiring
the SVI model and the SSVI surface parameterization (whence their
names).

\section{The star calibration
algorithm}\label{the-star-calibration-algorithm}

	\subsection{Anchored eSSVI slices with no Butterfly
arbitrage}\label{anchored-essvi-slices-with-no-butterfly-arbitrage}

	The key ingredient in our algorithm is a re-parameterization of a SSVI
slice, which constrains the slice to go through the data point
$(k^*, \theta^*)$ closest to the ATM (Forward), where $k$ denote the
log-forward moneyness and $\theta$ the total implied variance. Whence
the word \emph{anchored} in the section title. This re-parameterization
assumes that the data in this range are very reliable, which is
certainly true for not too-long term options on indexes at least.

So we change parameter: $\theta$ will be expressed in terms of the
parameters $\rho, \varphi$ and this new data-driven $(k^*, \theta^*)$
pair. At first order, solving $\theta^* = w(k^*, \theta)$ amounts simply
to $\theta = \theta^* - \rho \theta \varphi k^*$. We also substitute a
new parameter $\psi$ to the product $\theta \varphi$, so that eventually
our anchored smiles (anchored to $(k^*, \theta^*)$) are parameterized by
the pair $(\rho, \psi)$, $\theta$ being given by the formula
$\theta = \theta^* - \rho \psi k^*$.

This anchor trick can be seen as a refinement of Gatheral and Jacquier
initial idea to read the ATM Forward volatility on the market (and so,
to take it as a parameter): it avoids a pre processing step of the
market data which computes $\theta$ by interpolation from the available
bracketing strikes, which brings some noise, or the handling of $\theta$
as an additional parameter to calibrate, which adds a dimension.

Note that we could try to anchor to more than one point, yet this is
likely to put too many constraints on the parameters, especially for
large maturities.

	What is the allowed range for our new parameter $\psi$? Translating the
short term no butterfly constraint \eqref{eq: buttSuff2} reads:

\[\psi \leq 2\sqrt{\frac{\theta}{1+|\rho|}}\]

or yet $\psi^2 \leq \frac{4}{1+|\rho|}  (\theta^* - \rho \psi k^*)$
which is equivalent to the explicit bound

\[\psi \leq \psi_+(\rho, k^*, \theta^*)\]

where
$\psi_+(\rho, k^*, \theta^*)=\frac{-2 \rho k^*}{(1+|\rho|)}+\sqrt{\frac{4 \rho^2 (k^*)^2}{(1+|\rho|)^2} +\frac{4 \theta^*}{(1+|\rho|)}}$.

	In other words, \emph{all the no-butterfly arbitrage (in the sense that
they satisfy the Gatheral Jacquier bounds) eSSVI slices going through
the point $(k^*, \theta^*)$} (\emph{anchored} at $(k^*, \theta^*)$) are
parameterized by the SSVI formula, where $\theta$ is replaced by its
expression in terms of $(k^*, \theta^*)$, and the parameters $\rho$ an
$\psi$ are such that $\rho \in ]-1, 1[$ and
$0<\psi<\min{(\psi_+(\rho, k^*, \theta^*), \frac{4}{1+|\rho|})}$.

Also note that $\theta$ should be non negative, so that the constraint
$\psi < \frac{\theta^*}{\rho k^*}$ should be enforced when active.

	\subsection{Granting no Calendar-Spread arbitrage across
slices}\label{granting-no-calendar-spread-arbitrage-across-slices}

	Thanks to the result in Hendriks-Martini \cite{hendriks2017extended} we
have necessary and sufficient conditions for this. Let
$(\theta_i, \rho_i, \varphi_i)_{1 \leq i \leq N}$ a set of (e)SSVI slice
parameters corresponding to increasing time to maturities $0<T_1<..<T_N$
with $N>1$.

Then $\theta_i$ and $\psi_i$ should be non-decreasing, and the
condition:

\[\left|\frac{\rho_{i+1}\psi_{i+1}-\rho_{i}\psi_{i}}{\psi_{i+1}-\psi_i}\right| \leq 1\]

should hold.

	\subsection{Going forward calibration}\label{going-forward-calibration}

	Let us re-formulate those conditions in the setting where we calibrate
the slices \emph{going forward}: we start by calibrating
$(\theta_1, \rho_1, \varphi_1)$, so catering only for the absence of
Butterfly arbitrage for this initial slice.

The slices are then built in the following way, where we denote by
$\underline{\theta}, \underline{\psi}, \underline{\rho \psi}$ the
corresponding quantities associated to the \emph{previous} (already
calibrated) slice, and by $(\theta, \rho, \psi)$ the SSVI parameters for
the current slice.

The absence of Calendar Spread between the two slices is granted by the
conditions $\theta > \underline{\theta}$, $\psi > \underline{\psi}$ and
the last condition that reads
$-(\psi-\underline{\psi}) \leq \rho \psi -\underline{\rho \psi} \leq (\psi-\underline{\psi})$,
which amounts to $\psi \geq \psi_-(\rho)$ where

\[\psi_-(\rho):= \max{\left(\frac{\underline{\psi}-\underline{\rho \psi}}{1-\rho}, \frac{\underline{\psi}+\underline{\rho \psi}}{1+\rho}\right)}\]

	So we get again bound type conditions on $\psi$ given $\rho$ and the
previous slice parameters. Only the first condition
$\theta > \underline{\theta}$ is to be investigated: by substituting
$\theta = \theta^* - \rho \theta \varphi k^*$, it also amounts to a
bound type constraint $\psi>\hat{\psi}$ or $\psi<\hat{\psi}$ depending
on the sign of $\rho k^*$, where

\[\hat{\psi}:=\frac{\theta^*-\underline{\theta}}{\rho k^*}\]

Note that in the particular case $\rho=0$ one gets the constraint
$\theta^* > \underline{\theta}$, which is directly checked on the market
data at the current slice; this is indeed necessary as the smile is, in
this case, symmetrical and with a minimum for $k=0$, so that
$\theta^*>\theta$, which from the Calendar spread condition must be
bigger than $\underline \theta$.

\section{Implementation}\label{implementation}

	\subsection{Algorithm}\label{algorithm}

	Putting all the constraints together, for a given $\rho$, we get a set
of two-sided bound type constraints for $\psi$ (which is positive)
possibly empty, which grants simultaneously no Butterfly arbitrage and
no Calendar Spread arbitrage with the previous slice. Given any fit
objective function (a good choice is the $L^1$ norm of the price
differences between the eSSVI price and the market price, which has a
direct meaning from a financial point of view, since it is homogeneous
to a loss in monetary unit), we face for each slice a 2 dimensional
function in $\psi, \rho$. A very effective way to solve the minimization
problem is to proceed as follows:

\begin{enumerate}
\def\labelenumi{\arabic{enumi}.}
\itemsep1pt\parskip0pt\parsep0pt
\item
  Sample $\rho$ in the interval $]-1,1[$.
\item
  For each sampled $\rho$ use a Brent algorithm to find the point
  $\psi$, satisfying the constraints, at which the minimum of the
  objective function is obtained.
\item
  Pick up the minimum over all the $\rho$.
\item
  Repeat the procedure on a smaller interval centered on the optimal
  $\rho$ found before.
\end{enumerate}

This is very naive, yet very robust, quick and effective. Moreover the
minimization can be split $\rho$-wise on different cores. Note that one
could consider using a bi-dimensional minimizer here, yet it would not
be straightforward: indeed the domain is not a rectangular one; moreover
our experience is that the sampling in the correlation dimension can
remain crude given the smooth dependency of the prices with respect to
the correlation parameter. Moreover the parallelized one-dimension
approach grants to find a global minimum.

The global algorithm consists in calibrating the first slice, and then
the subsequent slices with the constraints produced by the calibrated
parameters attached to the previous slice.

	\subsection{Comments}\label{comments}

	\subsubsection{Choice of the initial
slice}\label{choice-of-the-initial-slice}

	It is natural to start from the short-term slice. Very often there is a
lot of curvature at short maturities (cf.~for instance Jim Gatheral
reference book \cite{gatheral2011volatility}), and the close-to-ATM
option price is roughly proportional to the ATM volatility, so that
there will be enough meaningful data to calibrate the initial pair
$\rho, \psi$. There might be issues though for very short maturities
where the market will convey only information on $\theta$ (or in this
case, equivalently, $\theta^\star$) and not on $\rho$ and $\varphi$, as
not ATM prices will carry little meaning, and consequently the analysis
from \cite{lee2005implied} can not be applied. Starting from the long
term end is more daring, since data is in general less reliable, and
there might be much less curvature due to the fact that implied
volatility smiles flatten for large maturities (see \cite{tehranchi} for
a theoretical study on the subject). Depending on the underlying (in
terms of liquidity) and the dataset (in terms of available
time-to-maturities), different strategies may be considered, including
intermediate ones where the initial slice is a mid term one and the
algorithm goes in both directions. In this case the algorithm should be
tweaked for the going-backward part, with computing the upper
constraints instead of the lower ones.

	Note that another idea is to calibrate $\rho$ from the ratio of the
slopes of the smile in the small strike and large strike regimes (cf.
\cite{de2009quasi} or \cite{GatheralJacquier}). This approach is not
easy to implement in practice though, because only few strikes are
available on the market at a given maturity.

	\subsubsection{Data consistency}\label{data-consistency}

	It may also happen that the feasible dataset for a given slice is empty
(though this never happened in our tests), due either to a too extreme
calibrated smile at a previous maturity or dubious data at the current
one. In particular if the $(k^\star, \theta^\star)$ data point is not
reliable, this algorithm should not be run.

	\subsubsection{Robustness}\label{robustness}

	This is the most appealing feature of this algorithm: besides the number
of sampling points of the correlation $\rho$ , there is no starting
point nor numerical parameter to set and tweak, the algorithm is
extremely robust and in this respect can be put safely in production. 20
points for the sampling of $\rho$ is enough according to us to grant a
calibration within the bid ask in general.

	\subsubsection{Stability of the calibrated
parameters}\label{stability-of-the-calibrated-parameters}

	At each slice, the free parameters correspond to the ATM Forward slope
and curvature. The fact that only 2 degrees of freedom remain after the
anchoring trick brings more stability than the traditional 3-dimensional
criterion. Lastly, the no Calendar Spread constraints ensure a built-in
consistency across slices which also brings a lot of stability to the
calibrated values.

	\subsubsection{Speed}\label{speed}

	Without any parallelization, a Python implementation for 12 maturities
and an average of 98 options per maturity (the number of options per
maturity varies between 68 and 184) takes $1.2$ seconds on a \emph{Intel
E5-2673 v3} processor.

On a more recent processor, as the \emph{Intel Xeon E7-8890 v3}, by
parallelizing the computation of the function $\rho$-wise the execution
time should be cut down to $0.1$ second or less. A C$\sharp$
implementation could also reduce the computation time by a factor 5
(quite a conservative estimate) to a final execution time of $0.01$
seconds or less.

 \section{Results}\label{results}

	We did numerous tests on several non public data sets by Zeliade
clients, on Equity Index and Equity Stock options. The algorithm
performed systematically very well, with a typical average option price
error below 4 bips of the underlying value. We display here some results
obtained on end-of-day SPX option quotes (data acquired from the CBOE,
https://datashop.cboe.com/option-quotes) on January 8th, 2018.

	\subsection{Data processing}\label{data-processing}

	We infer the Forward and Discount Factor at each available maturity by
robust linear regression leveraging the Put-Call-Parity for mid prices.
Then we select OTM options and filter out prices which are below 2 ticks
(the tick being 0.05 for SPX options). The rationale of this filtering
is that prices cannot be smaller than 1 tick, and rounding effects would
produce important distortions even for 2 ticks prices.

Implied volatility is computed using Jaeckel \emph{rational} algorithm
\cite{jackel2015let}.

    \begin{center}
    \adjustimage{max size={0.9\linewidth}{0.9\paperheight}}{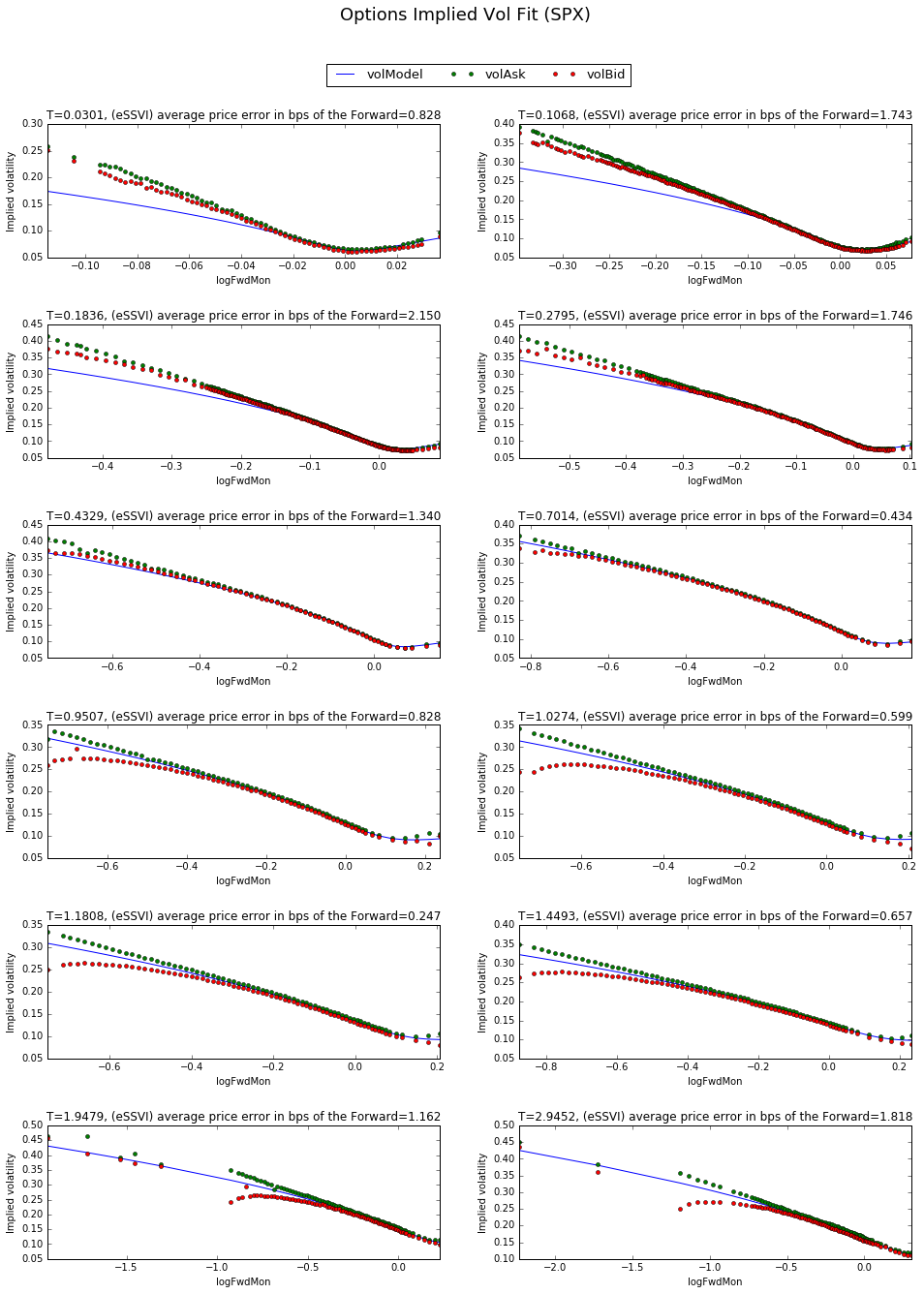}
    \end{center}
    { \hspace*{\fill} \\}

    \begin{center}
    \adjustimage{max size={0.9\linewidth}{0.9\paperheight}}{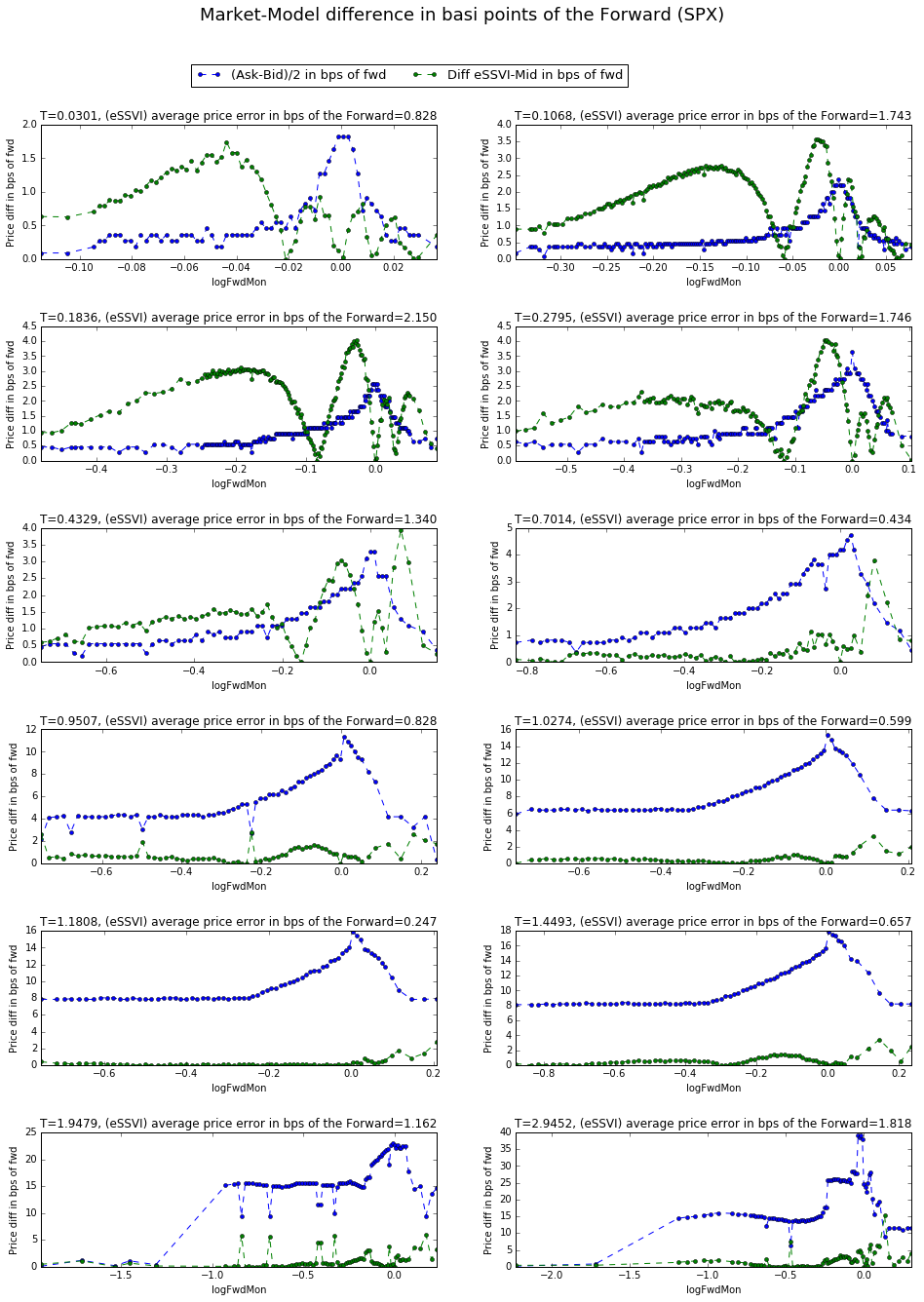}
    \end{center}
    { \hspace*{\fill} \\}
    
	In the figure above we plot in green the absolute error between the
calibrated price and the original Mid price, and in blue the Bid-Ask
spread halved, both expressed as basis points of the forward, i.e.~the
unit is $10^{-4} F_t$ where $F_t$ is the forward with maturity $t$ (the
same order of magnitudes would be obtained using the constant unit
$10^{-4} S_0$).

A good fit is thus identified when the green line is below the blue one.
Note that we think it is very important to look at the fit in implied
volatility scale \emph{and} in price scale altogether: indeed extreme
points in the surface in the short term range, or in the small and large
strike ranges, will have prices mostly given by the intrinsic value of
the option so that large error in implied volatility are not meaningful
to the extent that the price dos not depend significantly from the
implied volatility. The 2nd serie of plots allows to assess the price
error, especially in situations where the implied vol error is large.


\begin{table}
\centering   
\begin{tabular}{lrrrrr}
\toprule
{} &   Theta &    Psi &    Rho &  ATM vol (pct) &    Phi \\
Time to maturity &         &        &        &                &        \\
\midrule
0.030137         &  0.0001 &  0.012 & -0.224 &            6.4 &  96.33 \\
0.106849         &  0.0006 &  0.032 & -0.453 &            7.8 &  50.22 \\
0.183562         &  0.0014 &  0.049 & -0.495 &            8.6 &  35.82 \\
0.279452         &  0.0025 &  0.066 & -0.578 &            9.4 &  26.66 \\
0.432877         &  0.0049 &  0.089 & -0.610 &           10.6 &  18.38 \\
0.701370         &  0.0100 &  0.116 & -0.672 &           12.0 &  11.55 \\
0.950685         &  0.0158 &  0.131 & -0.704 &           12.9 &   8.28 \\
1.027397         &  0.0174 &  0.134 & -0.704 &           13.0 &   7.73 \\
1.180822         &  0.0215 &  0.145 & -0.725 &           13.5 &   6.75 \\
1.449315         &  0.0292 &  0.165 & -0.725 &           14.2 &   5.68 \\
1.947945         &  0.0444 &  0.191 & -0.746 &           15.1 &   4.29 \\
2.945205         &  0.0750 &  0.243 & -0.724 &           16.0 &   3.24 \\
\bottomrule
\end{tabular}
  \caption{Value of the calibrated parameters}
\end{table}

    \begin{center}
    \adjustimage{max size={0.9\linewidth}{0.9\paperheight}}{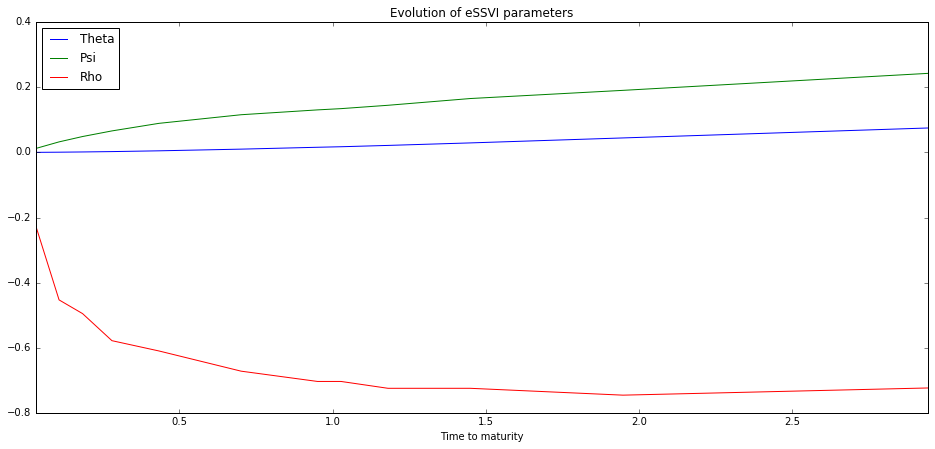}
    \end{center}
    { \hspace*{\fill} \\}
    
	\subsection{Comments}\label{comments}

	The implied vol fits are well within the Bid-Ask spread, except for the
3/4 shorter maturities for the left-end wing. Those visual vol
discrepancies translate though in very small errors for the price, as
confirmed by the price error plots, below 4 bips of the Forward (i.e.,
$4 \times 10^{-4} \times$ the Forward value). One should consider also
the traded volumes, which are almost zero in this range of strike for
those maturities.

Otherwise the fit in price is excellent across all the maturities.
Except for the very short maturities, the error we produce is in line,
or significatively smaller than the Bid-Ask spread. The shape of the
calibrated correlation is typical, and shows the benefit of eSSVI versus
the classical SSVI for calibrating simultaneously the medium/long
entries and the shorter ones. Lastly, the calibrated parameters evolve
smoothly with the time-to-maturity.

 \section{Arbitrage-free
interpolation}\label{arbitrage-free-interpolation}

	Our starting point in this section is a set of SSVI slice parameters
$(\theta_i, \psi_i, \rho_i)_{1 \leq i \leq n}$, attached to maturities
$0<T_1<..<T_n$ and such that:

\begin{itemize}
\itemsep1pt\parskip0pt\parsep0pt
\item
  Each slice is free of Butterfly arbitrage;
\item
  There is no Calendar Spread arbitrage between any 2 consecutive
  slices.
\end{itemize}

Note that the last property amounts to the fact that the total variance
smile attached to the longer maturity lies strictly above the smile
attached to the lower one. It follows from this geometrical point of
view that this property is transitive, so that there is no Calendar
Spread for all the slices globally.

	It is required in practice to get a continuous arbitrage free volatility
surface from these slices. We show below that the natural interpolation
and extrapolation of the eSSVI parameters provides a
\textbf{\emph{continuous eSSVI surface}} which is indeed arbitrage free.
This is a very nice property of the eSSVI parameterization. Note that it
is by no way built-in or automatic.

	\subsection{Interpolation scheme}\label{interpolation-scheme}

	We describe the interpolation scheme between 2 consecutive slices, which
we denote by $(\theta_i, \psi_i, \rho_i)$ and
$(\theta_{i+1}, \psi_{i+1}, \rho_{i+1})$.

	The no arbitrage conditions read:

\begin{enumerate}
\def\labelenumi{\arabic{enumi}.}
\itemsep1pt\parskip0pt\parsep0pt
\item
  $\theta_{i+1}>\theta_i$
\item
  $\psi_{i+1} \geq \psi_i$
\item
  $\psi_j \leq \min\left( \frac{4}{1+|\rho_j|}, 2 \sqrt{\frac{\theta_j}{1+|\rho_j|}} \right)$
  for $j =i,i+1$
\item
  $\left| \frac{\rho_{i+1}\psi_{i+1}-\rho_i\psi_i}{\psi_{i+1}-\psi_i} \right| \leq 1$
\end{enumerate}

	For $\lambda \in [0,1]$ we define the following interpolation scheme:

\begin{itemize}
\itemsep1pt\parskip0pt\parsep0pt
\item
  $\theta_\lambda = (1-\lambda) \theta_i + \lambda\theta_{i+1}$;
\item
  $\psi _\lambda =  (1-\lambda) \psi_i + \lambda\psi_{i+1}$;
\item
  $\rho_\lambda \psi _\lambda =  (1-\lambda) \rho_i\psi_i + \lambda\rho_{i+1}\psi_{i+1}$.
\end{itemize}

Each such slice will be attached to a maturity $t$ such that
$\lambda = \frac{t-T_i}{T_{i+1}-T_i}$.

	\subsubsection{Calendar Spread
arbitrage}\label{calendar-spread-arbitrage}

	Since $\theta_\lambda$ and $\psi_\lambda$ interpolate linearly between
ordered quantities, we will have $\theta_\lambda < \theta_\mu$ and
$\psi_\lambda < \psi_\mu$ for $0 \leq \lambda < \mu \leq  1$. In the
same way since
$\rho_\lambda \psi_\lambda -\rho_\mu \psi_\mu = (\lambda-\mu) (\rho_{i+1}\psi_{i+1}-\rho_i\psi_i)$
and $\psi_\lambda - \psi_\mu = (\lambda-\mu) (\psi_{i+1}-\psi_i)$
condition 4 is satisfied also. So there is no calendar spread arbitrage
in between 2 interpolated slices within the same bucket $(i; i+1)$.

	By the transitivity property above we deduce that there is no arbitrage
between 2 interpolated slices in different buckets.

	\subsubsection{Butterfly arbitrage}\label{butterfly-arbitrage}

	To alleviate notations we will write the proof for the 1st maturity
bucket.

	We start by checking that $\psi_\lambda<\frac{4}{1+|\rho_\lambda|}$
holds. Since the condition is verified for $\lambda = 0,1$ it suffices
to show that the derivative of
$f(\lambda) = \psi_{\lambda}(1+|\rho_{\lambda}|) = \psi_{\lambda}+|\rho_{\lambda}\psi_{\lambda}|$
has constant sign. Since
$f'(\lambda) = \psi_2-\psi_1 + (\rho_2 \psi_2- \rho_1 \psi_1) \text{sign}(\rho_{\lambda})$
and $\psi_{\lambda}, \rho_{\lambda}$ satisfy condition \emph{2.} and
\emph{4.}, we conclude that $f'>0$.

	Moreover, whenever $\rho_1$ and $\rho_2$ have the same sign, the
function $f$ is linear, while if the sign of $\rho$ changes, the
function $f$ is piecewise linear.

	We now check that the condition
$\psi_\lambda< 2\sqrt{\frac{\theta_\lambda}{1+|\rho_\lambda|}}$ holds.
This is equivalent to requiring that \[
\psi_\lambda(\psi_\lambda+|\rho_\lambda \psi_\lambda|)< 4 \theta_\lambda
\]

	We will start by considering the case where $\rho_1$ and $\rho_2$ have
the same sign:

In this case we can rewrite the previous equation as \[
\left(\psi_1 + \lambda(\psi_2-\psi_1)\right)(a+b\lambda)< 4 \left(\theta_1 + \lambda(\theta_2-\theta_1)\right)\,,
\] where \[
a = \psi_1(1+|\rho_1|)\quad\text{and}\quad b = \psi_2-\psi_1+|\rho_2|\psi_2- |\rho_1|\psi_1 >0\,.
\]

	Observe that the LHS is a convex function since $b>0$, so it lies below
its chord on $[0,1]$, which in turn lies below the RHS since the
requirement is fulfilled for $\lambda = 0,1$.

	We are now left with the case where $\rho_0$ and $\rho_1$ have different
signs:

In this case there is a unique $\lambda^*$ such that
$\rho_{\lambda^*} = 0$. Condition $3.$ reduces, for $\lambda^*$, to \[
\psi_{\lambda^*} <\min(4, 2\sqrt{\theta_{\lambda^*}})\,.
\]

Since $\psi_i< 4$, $\psi_i< 2 \sqrt{\theta_i}$ $i = 0,1$, and since the
square root is a concave function, the condition above is satisfied at
$\lambda^*$.

Since $f$ is linear on each interval $[0,\lambda^*]$ and
$[\lambda^*, 1]$ we can apply the reasoning for $\rho$ with constant
sign on the two intervals $[0,\lambda^*]$ and $[\lambda^*, 1]$ to
conclude that the no-arbitrage conditions are verified also in the case
in which $\rho_0$ and $\rho_1$ have different signs.

	\subsection{Short term extrapolation}\label{short-term-extrapolation}

	How to extrapolate to the time bucket $]0, T_1[$?

	Observe first that it is natural to require that the option price
$C(t, k)$ has the minimal continuity property that
$C(t,0) \to (S_0-S_0)_+=0$ as $t \to 0+$; this is not required by the no
arbitrage theory, but we look for \emph{continuous} continuous-time
formulas for option prices. For small maturities $t$, the ATM Black $\&$
Scholes formula can be approximated by
$C(t, 0) \approx S_0\left(1-2\Phi\left( \frac{\sqrt{\theta_t}}{2}\right)\right)$,
where $\Phi$ denotes the Gaussian cumulated density function, so that
this continuity statement is equivalent to the property that
$\theta_t \to 0$.

	In our eSSVI parametrization, the no arbitrage condition 3. implies that
$\psi_t$ goes to zero as well. Therefore the simplest short term
extrapolation scheme is as follows:

	\begin{itemize}
\itemsep1pt\parskip0pt\parsep0pt
\item
  $\theta_t = \lambda\theta_1$;
\item
  $\psi _t =  \lambda\psi_1$;
\item
  $\rho_t = \rho_1$.
\end{itemize}

Here $\lambda = \frac{t}{T_1}$.

	Then conditions 1., 2. and 4. are readily checked. Conditions 3. reads
in turn:

\[ \lambda \psi_1< \min\left( \frac{4}{1+|\rho_1|}, 2 \sqrt{\frac{\lambda \theta_1}{1+|\rho_1|}} \right)\]

Now $\lambda \psi_1< \psi_1 < \frac{4}{1+|\rho_1|}$ and
$\lambda \psi_1< 2 \sqrt{\frac{\lambda \theta_1}{1+|\rho_1|}}$ follows
from the fact that
$\sqrt{\lambda} \psi_1 < \psi_1 < 2 \sqrt{\frac{ \theta_1}{1+|\rho_1|}}$
for $\lambda<1$.

	Lastly the absence of Calendar Spread arbitrage between 2 slices within
the first maturity bucket is shown exactly as above, and between a slice
in the first maturity bucket and another one after $T_1$ by
transitivity.

	\subsection{Long term extrapolation}\label{long-term-extrapolation}

	To extrapolate beyond $T_N$, pick up any continuous increasing function
$u(t)$ on $[T_N, \infty[$ such that $u(T_N)=0$, and set:

\begin{itemize}
\itemsep1pt\parskip0pt\parsep0pt
\item
  $\theta_t = \theta_N+u(t)$;
\item
  $\psi_t = \psi_N$;
\item
  $\rho_t = \rho_N$.
\end{itemize}

Then conditions 1. to 4. are readily checked. In the same way there is
no calendar spread arbitrage between 2 slices living beyond $T_N$. With
the same transitivity argument as before, there is no calendar spread
arbitrage between one such slice and a slice living below $T_N$.

\section{Conclusion}\label{conclusion}

	We have designed a novel calibration algorithm of the eSSVI model, which
relies on the forward slice-by-slice calibration of SSVI slices
constrained to go exactly through the data point closest to the Forward
of each maturity, computing explicitly the no Butterfly and no Calendar
Spread constraints. The naive piecewise interpolation/extrapolation of
the slice parameters is shown to be also free of arbitrage. All in all
we have a simple, quick and robust calibration algorithm of the
volatility surface, which fits very well except maybe in the more
demanding (tight market-making) situations. Moreover it is
straightforward to \emph{store} and \emph{re-use} the calibrated
parameters $(\theta_i, \rho_i, \psi_i)_{1 < i < N}$ alongside the market
parameters $(T_i, F_i, DF_i)_{1 < i < N}$ (where $F$ denoted the Forward
and $DF$ the Discount Factor) to parsimoniously serialize the whole
volatility surface, which is very useful for constituting
\emph{histories} of volatility surfaces, e.g.~for risk purposes.


    \end{document}